\documentclass{ws-ijmpe}
\usepackage[super,compress]{cite}

\newcommand{\MeV}{{\rm \,MeV}}

\newcommand{\HQSS}{{\rm HQSS}}
\newcommand{\WT}{{\rm WT}}
\newcommand{\SU}{\mbox{SU}}
\newcommand{\U}{\mbox{U}}
\newcommand{\bc}{{\bar{c}}}
\newcommand{\bq}{{\bar{q}}}

\newcommand{\bQ}{\bar{Q}}

\newcommand{\ben}{\begin{enumerate}}
\newcommand{\een}{\end{enumerate}}

\newcommand{\be}{\begin{equation}}
\newcommand{\ee}{\end{equation}}
\newcommand{\bea}{\begin{eqnarray}}
\newcommand{\eea}{\end{eqnarray}}
\newcommand{\ds}{\begin{displaystyle}}
\renewcommand{\ss}{\begin{scriptstyle}}

\newcommand{\Eq}[1]{Eq.~(\ref{eq:#1})}

\newcommand{\ignore}[1]{}

\newcommand{\ba}{\begin{eqnarray}}
\newcommand{\ea}{\end{eqnarray}}

\begin{document}

\markboth{Tolos}{Charming mesons with baryons and nuclei}

\catchline{}{}{}{}{}

\title{Charming mesons with baryons and  nuclei}

\author{\footnotesize Laura Tolos$^{1,2}$ \footnote{e-mail: tolos@ice.csic.es}}

\address{
$^1$Instituto de Ciencias del Espacio (IEEC/CSIC), Campus Universitat 
Aut\`onoma de Barcelona, Facultat de Ci\`encies, Torre C5, E-08193 Bellaterra 
(Barcelona), Spain\\
$^2$Frankfurt Institute for Advanced Studies, Johann Wolfgang Goethe University, Ruth-Moufang-Str. 1,
60438 Frankfurt am Main, Germany
}

\maketitle

\begin{history}
\received{Day Month Year}
\revised{Day Month Year}
\end{history}

\begin{abstract}
The properties of charmed mesons in nuclear matter and nuclei are reviewed. Different frameworks are discussed paying a special attention to unitarized coupled-channel approaches which incorporate heavy-quark spin symmetry. Several charmed baryon states with negative parity are generated dynamically by the $s$-wave interaction between pseudoscalar and vector meson multiplets with $1/2^+$ and $3/2^+$ baryons. These states are compared to experimental data. Moreover, the properties of open-charm mesons in matter are analyzed.   The in-medium solution accounts for Pauli blocking effects, and for the meson self-energies in a self-consistent manner. The behavior in the nuclear medium of the rich spectrum of dynamically-generated baryon states is studied as well as their influence in the
self-energy and, hence, the spectral function of open charm. The possible experimental signatures of the in-medium properties of open charm are finally  addressed, such as the formation of charmed nuclei, in connection with the future FAIR facility.

\end{abstract}


\ccode{PACS numbers: 14.20.Lq; 14.40.Lb;  21.65.Jk; 21.85.+d}


\section{Introduction}
\label{intro}

Quantum Chromodynamics (QCD) is considered to be the basic theory of the strong interaction. In the low-energy regime, QCD becomes a strongly-coupled theory, many aspects of which are not yet understood. A particular effort has been invested in exploring the QCD phase diagram for high density and/or temperatures. It is of great interest to understand the mechanisms inducing the transition from confined quarks inside hadrons to a deconfined plasma made of quarks and gluons as well as the restoration of certain QCD symmetries, such as the chiral symmetry in the light-quark sector, under extreme conditions of density and temperature  Thus, the study of matter under extreme conditions has become one of the main research activities of several experimental programs, from the ongoing LHC/CERN project (Switzerland) \cite{lhc} to the forthcoming PANDA and CBM experiments at FAIR (Germany) \cite{fair}. Until now, due to the restrictions in the available energy, the studies have been concentrated in matter within the light-quark sector. With the on-going and upcoming research facilities, the aim is also to move from the light-quark domain to the heavy-quark one and to face new challenges where charm and new symmetries, such as heavy-quark symmetry, will play a dominant role.

Interest in the properties of open and hidden charmed mesons was triggered more than 20 years ago in the context of relativistic nucleus-nucleus collisions in connection with charmonium suppression \cite{Matsui:1986dk}  as a probe for the formation of quark-gluon plasma (QGP). Nowadays, the nature of newly observed baryon and meson states with the charm degree of freedom is a matter of high interest in connection with many experiments, such as CLEO, Belle, BABAR \cite{cleo,belle,babar} and others. The goal is to understand whether these states can be accommodated within the quark model picture and/or qualify better as being dynamically generated via hadron-hadron scattering processes. To this end, a large part of the experimental program in hadronic physics at PANDA (FAIR)  will be devoted to charmonium spectroscopy. Also, the CBM (FAIR) experiment will extend the GSI program for in-medium modification of hadrons in the light quark sector and provide the first insight into the charm-nucleus interaction. Indeed, the influence of medium modifications in the charmonium production at finite baryon densities would affect the formation of the QGP phase of QCD at high densities.

This review aims at overviewing the recent advances on the properties of charmed hadrons in dense matter and nuclei. We first start by analyzing baryon states with charm degrees of freedom in connection to experimental results. These states are generated dynamically within the context of unitarized meson-baryon coupled-channel models. We will pay a special attention to those which implement heavy-quark spin symmetry, which is a proper QCD symmetry in the limit of infinite quark masses. Then we implement medium corrections to the meson-baryon interaction in order to address  the in-medium properties of open-charm mesons within self-consistent unitarized coupled-channel models. Finally, the  formation of charmed mesic nuclei is considered  as a possible experimental scenario to test charmed meson properties in matter at FAIR.

\section{Unitarized meson-baryon coupled-channel models with charm}

One of the primary goals in the physics of hadrons is to understand the nature of newly discovered states, whether they can be described within the quark model picture and/or as hadron-hadron molecules. In this section we will adopt the latter approach and analyze the recent developments in the description of experimental baryon states which incorporate the charm degree of freedom. These states are dynamically generated by means of the scattering of mesons and baryons in a unitarized coupled-channel description.  It is, however, known that some baryon states can be constructed as $qqq$ entity in a quark model, and simultaneously as a dynamically generated state in a meson-baryon coupled-channel description (that is a $qqq -q \bar q $ molecular state) \cite{Klempt:2009pi}, though, some of their properties might differ. 

Given the success of unitarized coupled-channel approaches in the description of some of the existing experimental data in the light-quark sector,  the charm degree of freedom has been recently incorporated in these models \cite{Tolos:2004yg, Tolos:2005ft, Lutz:2003jw, Lutz:2005ip, Hofmann:2005sw, Hofmann:2006qx, Lutz:2005vx, Mizutani:2006vq, Tolos:2007vh,  JimenezTejero:2009vq, Haidenbauer:2007jq, Haidenbauer:2008ff,  Haidenbauer:2010ch, Wu:2010jy, Wu:2010vk, Wu:2012md, Oset:2012ap} and several experimental states have been described as dynamically-generated baryon molecules.  This is the case, for example, of the $\Lambda_c(2595)$, which is the charm sector counterpart of the $\Lambda(1405)$. 

Whereas a separable potential for the bare meson-baryon interaction with only up, down and charm degrees of freedom was assumed in Ref.~\cite{Tolos:2004yg}, some of these approaches are based on a bare meson-baryon interaction saturated with the $t$-channel exchange of vector mesons between pseudoscalar mesons and baryons in the zero-range approximation while preserving chiral symmetry for light mesons \cite{Lutz:2003jw, Lutz:2005ip, Hofmann:2005sw,Hofmann:2006qx,Mizutani:2006vq}. Later works revisited the zero-range approach by using the full $t$-dependence of the $t$-channel vector-exchange driving term \cite{JimenezTejero:2009vq}.  Other approaches have made use of the J\"ulich meson-exchange model \cite{Haidenbauer:2007jq, Haidenbauer:2008ff, Haidenbauer:2010ch} by deriving the transition potential in close analogy to the meson-exchange $ \bar K N$ model while using, as a working hypothesis, SU(4) symmetry constraints. And some others have relied on the hidden gauge formalism \cite{Wu:2010jy, Wu:2010vk, Wu:2012md, Oset:2012ap} to describe dynamically baryon states found experimentally.

All these models, however, do not explicitly incorporate heavy-quark spin symmetry (HQSS)~\cite{Isgur:1989vq,Neubert:1993mb,Manohar:2000dt} and, thus, it is unclear whether  they fulfilled the constraints imposed by HQSS. \footnote{A detailed analysis of the hidden-gauge models of Refs.~\cite{Wu:2010jy,Wu:2010vk,Wu:2010rv} shows no actual violation in the heavy-quark limit (see Ref.~\cite{Xiao:2013yca}).} HQSS is a QCD symmetry that appears when the quark masses, such as the charm mass, become larger than the typical confinement scale. In the following, we analyze the implementation of HQSS in meson-baryon interactions as the charm degree of freedom is incorporated.

\subsection{Extended Weinberg-Tomozawa to spin-flavor with heavy-quark spin symmetry constraints}
\label{su8wt}

HQSS predicts that all types of spin interactions involving heavy quarks vanish for infinitely massive quarks. Thus, HQSS connects vector and pseudoscalar mesons containing charmed quarks. On the other hand, chiral symmetry fixes the lowest order interaction between Goldstone bosons and other hadrons in a model independent way; this is
the Weinberg-Tomozawa (WT) interaction. Thus, it is appealing to have a predictive model for four flavors including all basic hadrons (pseudoscalar and vector mesons, and $1/2^+$ and $3/2^+$ baryons) which reduces to the WT interaction in the sector where Goldstone bosons are involved and which incorporates HQSS in the sector where charm quarks participate. This model was developed in Refs.~\cite{GarciaRecio:2008dp, Gamermann:2010zz, Romanets:2012hm,Garcia-Recio:2013gaa}, as an extension of the SU(6) approach in the light sector of Refs.~\cite{GarciaRecio:2005hy, GarciaRecio:2006wb, Toki:2007ab, GarciaRecio:2010ki,Gamermann:2011mq,Garcia-Recio:2013uva}. Note that this  model can be used not only for meson-baryon systems with charm content but also in the bottom sector \cite{GarciaRecio:2012db}, since we expect HQSS to be better realized in heavier systems. Here we reproduce the main features.

We start with the extension of the WT meson-baryon interaction to spin-flavor symmetry (SF) with HQSS constraints. The extension for the on-shell vertex is \cite{GarciaRecio:2005hy}
\begin{equation}
V_\WT^{\rm sf} = \frac{K(s)}{4 f^2} 4 J_M^i J_B^i,
\qquad
i=1,\ldots,(2N_F)^2-1,
\label{eq:2.8}
\end{equation}
where $K(s)$ is a function that depends on the meson-baryon energy, and $J_M^i$ and $J_B^i$ are the $\SU(2N_F)$ generators on mesons and
baryons.  Mesons consists of $0^-$ $(P)$ and $1^-$ $(V)$ lowest-lying states,
while baryons contain $\frac{1}{2}^+$ $(B)$ and $\frac{3}{2}^+$ $(B^*)$
lowest-lying states. When this interaction is restricted to the sector $PB\to
PB$, it reproduces the standard WT off $B$ targets. 

The corresponding Hamiltonian for number of flavors $N_F$ and three colors 
reads~\cite{GarciaRecio:2006wb}
\begin{equation}
{\mathcal H}^{\rm sf}_{\rm WT}(x)
= -\frac{{\rm i}}{4f^2} :[\Phi, \partial_0 \Phi]^A{}_B
{\cal B}^\dagger_{ACD} {\cal B}^{BCD}:
,
\quad
A,B,\ldots = 1,\ldots,2N_F
,
\label{eq:2.9}
\end{equation}
where $\Phi^A{}_B(x)$ is the meson field,  
which contains the fields of $0^-$ (pseudoscalar) and $1^-$ (vector)
mesons, and ${\cal B}^{ABC}$ is a completely symmetric
tensor, which contains the fields of the lowest-lying baryons
with $J^P=\frac{1}{2}^+$ and $\frac{3}{2}^+$.

\begin{figure}
\begin{center}
\includegraphics[width=0.7\textwidth]{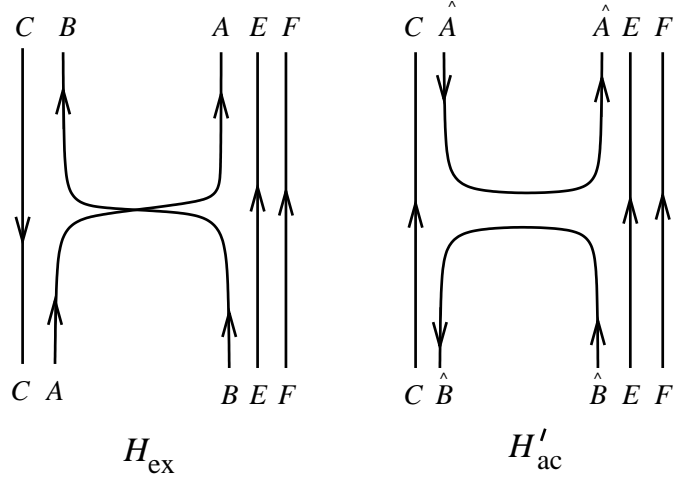}
\caption{\small The two mechanisms acting in the spin-flavor extended WT
  interaction. $H_{\rm ex}$ (exchange of quarks) and $H^{\prime}_{\rm ac}$
  (annihilation and creation of quark-antiquark pairs) corrected by HQSS constraints. The indices $\hat{A}$ and $\hat{B}$ are restricted to light quarks.}
\label{fig:hexhac}
\end{center}
\end{figure}

Extracting the kinematical part in \Eq{2.8}, let
\begin{equation}
H_\WT = 4 J_M^i J_B^i
.
\end{equation}
This operator can be written in terms of meson and baryon operators
\cite{GarciaRecio:2006wb,GarciaRecio:2008dp}, and it contains two distinct
mechanisms which stem from expanding the meson commutator in \Eq{2.9},
\begin{eqnarray}
H_\WT &=& H_{\rm ex} + H_{\rm ac}
,
\nonumber \\
H_{\rm ex} &=&
:M^A{}_C M^{\dagger C}{}_B B^{BDE} B^\dagger_{ADE}:
,
\nonumber \\
H_{\rm ac} &=& 
  -: M^\dagger{}^A{}_C M^C{}_B B^{BDE} B^\dagger_{ADE}:
,
\quad
A,\ldots,E = 1,\ldots,2N_F
.
\end{eqnarray}
Here $M^A{}_B$ and $B^{ABC}$ are the annihilation operators of mesons and
baryons, respectively, with $M^\dagger{}^A{}_B = (M^B{}_A)^\dagger$, and
$B^\dagger_{ABC}=(B^{ABC})^\dagger$. $B^{ABC}$ is a completely symmetric
tensor. They are normalized as
\begin{eqnarray}
[ M^A{}_B, M^{\dagger C}{}_D] &=& \delta^A_D\delta^C_B
,\nonumber \\
\{ B^{ABC}, B^\dagger_{{A^\prime}{B^\prime}{C^\prime}} \}
&=&
\delta^A_{A^\prime} \delta^B_{B^\prime} \delta^C_{C^\prime}
+ \cdots \mbox{~~(6 permutations)}
.
\label{eq:2.11}
\end{eqnarray}
Schematically, representing the quark and antiquark operators by $Q^A$ and
$\bQ_A$,
\begin{equation}
M^A{}_B\sim Q^A\bQ_B, \quad M^\dagger{}^A{}_B\sim \bQ^\dagger{}^A
Q^\dagger_B,\quad B^{ABC}\sim Q^A Q^B Q^C, \quad B^\dagger_{ABC}\sim Q^\dagger_A
Q^\dagger_B Q^\dagger_C \ .
\end{equation}
So, an upper index in $M$ or $B$ represents the SF of a quark to be
annihilated, whereas in $M^\dagger$ it represents that of an antiquark to be
created.  Likewise, a lower index in $M^\dagger$ or $B^\dagger$ represents the
SF of a quark to be created while in $M$ it represents that of an antiquark to
be annihilated. From this identification it is immediate to interpret the two
mechanisms $H_{\rm ex}$ and $H_{\rm ac}$ in terms of quark and antiquark
propagation: the exchange part $H_{\rm ex}$, in which a quark is
transferred from the meson to the baryon, as another one is transferred from
the baryon to the meson; and the annihilation-creation $H_{\rm ac}$  mechanism, where
an antiquark in the meson annihilates with a similar quark
in the baryon, with subsequent creation of a quark and an antiquark.
The $H_{\rm ac}$ can violate  HQSS when the annihilation or
creation of $q\bq$ pairs involve heavy quarks, as in the heavy-quark limit the number of charm quarks and the number of charm 
antiquarks are separately conserved (implying $\U_c(1)\times \U_\bc(1)$). Indeed, the HQSS group, which also includes a group of separate rotations of the $c$ quark and 
$\bar c$ antiquark, reads as ${\rm HQSS}=\SU_c(2) \times \SU_\bc(2) \times \U_c(1)\times \U_\bc(1)$.  A simple solution to enforce
HQSS with minimal modifications is to remove just the offending contributions in $H_{\rm ac}$, which come from creation or annihilation of charm
quark-antiquark pairs. This implies to remove the interaction when the labels
$A$ or $B$ are of heavy type in $H_{\rm ac}$ to construct $H^{\prime}_{\rm ac}$ as
\begin{equation}
H^\prime_{\rm ac} =
  -: M^\dagger{}^{\hat{A}}{}_C M^C{}_{\hat{B}} B^{{\hat{B}}DE} B^\dagger_{{\hat{A}}DE}:
,
\quad
C,D,E = 1,\ldots,8
,\quad
\hat{A},\hat{B} = 1,\ldots,6
.
\label{eq:2.21a}
\end{equation}
The indices with hat are restricted to $\SU(6)$. The two final mechanisms that respect HQSS, $H_{\rm ex}$ and $H^{\prime}_{\rm ac}$ are shown in Fig.~\ref{fig:hexhac}.
 
 To summarize, the extended WT model to SF with HQSS constraints is given by
\begin{equation}
V = \frac{K(s)}{4f^2} H^\prime_\WT
,\qquad
H^\prime_\WT = H_{\rm ex}+H^\prime_{\rm ac}
.
\label{eq:2.21}
\end{equation}
This model fulfills some desirable requirements: i) it has symmetry
$\SU(6)\times \HQSS$, i.e., SF symmetry in the light sector and HQSS in the
heavy sector, the two invariances being compatible; ii) it reduces to SU(6)-WT
in the light sector, so it is consistent with chiral symmetry in that
sector. 

\subsection{Coupled-channel unitarization and symmetry breaking}

The final expression to be used for the extended WT interaction in the coupled meson-baryon basis for total charm $C$, strangeness $S$,  isospin $I$ and spin $J$ is
\begin{equation}
V_{ij}^{CSIJ} =
D_{ij}^{CSIJ}\,\frac{2\sqrt{s}-M_i-M_j}{4f_if_j} 
\sqrt{\frac{E_i+M_i}{2M_i}}\sqrt{\frac{E_j+M_j}{2M_j}}, 
\label{eq:pot}
\end{equation}
where $\sqrt{s}$ is the center of mass (C.M.) energy of the system; $E_i$ and
$M_i$ are, respectively, the C.M. energy and mass of the baryon in the channel
$i$; and $f_i$ is the decay constant of the meson in the $i$-channel.  The hadron masses
and meson decay constants can be found in
Ref.~\cite{Romanets:2012hm}  \footnote{Other alternative expression for the WT can be found in Ref.~\cite{Garcia-Recio:2013gaa}. In this work, 
the $(2\sqrt{s}-M_i-M_j)$ term, with $M_i$ and $M_j$ the incoming
  and outgoing baryon masses, has been approximated by the sum of the CM energies of the incoming
  and outgoing mesons. This non-relativistic approximation for the baryons is consistent with the
  treatment for the baryons to implement the
  HQSS constraints. Moreover, the $\sqrt{(E+M)/(2M)}$ factors
  included in the potential have been moved to the definition
  of the loop function.}.

In order to calculate the scattering amplitudes, $T_{ij}$, we solve the
on-shell Bethe-Salpeter equation (BSE),  using the matrix $V^{CSIJ}$ as kernel:
\begin{equation}
T^{CSIJ}=(1-V^{CSIJ}G^{CSIJ})^{-1}V^{CSIJ}\label{eq:bse} \ ,
\end{equation}
where $G^{CSIJ}$ is a diagonal matrix containing the meson-baryon propagator
in each channel.  Explicitly,
\begin{equation}
G^{CSIJ}_{ii} = 2M_i
\left(
\bar{J}_0(\sqrt{s};M_i,m_i) - \bar{J}_0(\mu^{SI};M_i,m_i)
\right)
, 
\label{eq:loop}
\end{equation}
with $M_i$ ($m_i$) the mass of the baryon (meson) in the channel $i$. The loop
function $\bar{J}_0$ can be found in the appendix of Ref.~\cite{Nieves:2001wt}
(Eq.~A9) for the different possible Riemann sheets.  The meson-baryon
propagator (loop) is logarithmically ultraviolet divergent, thus, it needs to
be renormalized. This has been done by a subtraction point regularization such
that
\begin{equation}
G_{ii}^{CSIJ} (\sqrt{s})=0 \quad\text{at~~} \sqrt{s}=\mu^{CSI},
\label{eq:musi}
\end{equation}
with $\mu^{CSI} =\sqrt{m_{\rm{th}}^2+M_{\rm{th}}^2}$, where $m_{\rm{th}}$ and
$M_{\rm{th}}$, are, respectively, the masses of the meson and baryon producing
the lowest threshold (minimal value of $m_{\rm{th}}+M_{\rm{th}}$) for each
$CSI$ sector, independent of the angular momentum $J$.  This renormalization
scheme was first proposed in Refs.~\cite{Hofmann:2005sw,Hofmann:2006qx} and it
was successfully used in Refs.~\cite{GarciaRecio:2003ks,GarciaRecio:2008dp,Romanets:2012hm,Garcia-Recio:2013gaa,GarciaRecio:2012db}. A recent
discussion on the regularization method can be found in
Ref.~\cite{Hyodo:2008xr}.

The dynamically-generated baryon states appear as poles of the scattering
amplitudes on the complex energy $\sqrt{s}$ plane. The poles of the scattering amplitude on the first
Riemann sheet that appear on the real axis below threshold are interpreted as
{\it bound states}. The poles that are found on the second Riemann sheet below the
real axis and above threshold are identified with {\it resonances}.  The mass and the width of the state can be found from
the position of the pole on the complex energy plane. Close to the pole, the
scattering amplitude behaves as
\begin{equation} \label{Tfit} 
T^{CSIJ}_{ij} (s) \approx \frac{g_i
e^{i\phi_i}\,g_je^{i\phi_j}}{\sqrt{s}-\sqrt{s_R}} \,.  
\end{equation} %
The mass $M_R$ and width $\Gamma_R$ of the state result from
$\sqrt{s_R}=M_R - \rm{i}\, \Gamma_R/2$, while $g_j e^{i\phi_j}$ (modulus and
phase) is the coupling of the state to the $j$-channel.

The dynamically-generated states can be classified under the symmetry group $\SU(6)\times\HQSS$. Indeed, dynamically generated states are most likely to occur coming from the most attractive $\SU(6)\times\HQSS$ representations. However, the $\SU(6)\times\HQSS$ is strongly broken in nature and several soft symmetry-breaking mechanisms have to be introduced in order to compare the dynamically-generated states to the experimentally observed ones. This is performed by an adiabatic change of hadron masses and meson weak-decay constants. We consider the breaking of the light SF SU(6) to $\SU(3) \times \SU_{J_l}(2)$, with $J_l$ the spin of the light quarks.  Subsequently, we break the SU(3) light flavor group to SU(2) isospin symmetry group, preserving the HQSS.  Thus, we assume exact isospin, total spin and flavor conservation. In this way we can assign SU(3) and SU(6) representation labels to each found state, and also identify the HQSS multiplets. 

\section{Dynamically-generated baryon states with charm content}
\label{dyn}

\begin{table}
\begin{center}
\begin{tabular}{lccccc }\hline
Resonance & $I (J^P)$ & Status & Mass [MeV] & $\Gamma$ [MeV] & Prediction \\
& & & & & Mass [MeV] ($\Gamma$[MeV])
\\
\hline
$\Lambda_c(2595)$ &  $0(1/2^-)$ & *** & $2592.25 \pm 0.28$ & $2.6 \pm 0.6$ & $2618.8 \, (1.2)$  \\
$\Lambda_c(2625)$ &  $0(?^?)$ & *** & $2628.11 \pm 0.19$ & $<0.97$ &  $2666.6 \,(53.7)$ \\
$\Lambda_c(2765)$ &  $?(?^?)$ & * & $2766.6 \pm 2.4$ & $50$ \\
or $\Sigma_c(2765)$ & & & & \\
$\Lambda_c(2880)$ & $0(5/2^+)$ & *** & $2881.53 \pm 0.35$ & $5.8 \pm 1.1$\\
$\Lambda_c(2940)$ & $0(?^?)$ & *** & $2939.8^{+1.4}_{-1.5}$ & $17^{+8}_{-6}$  \\
$\Sigma_c(2800)^{++}$ & $1(?^?)$ & *** & $2801^{+4}_{-6}$ & $75^{+22}_{-17}$ \\
$\Sigma_c(2800)^{+}$ & $1(?^?)$ & *** & $2792^{+14}_{-5}$ & $62^{+60}_{-40}$ \\
$\Sigma_c(2800)^{0}$ & $1(?^?)$ & *** & $2806^{+5}_{-7}$ & $72^{+22}_{-15}$ \\
$\Xi_c(2790)^{+}$ & $1/2(?^?)$ &*** & $2789.1 \pm 3.2$ & $ < 15$ & $2804.8 \, (20.7)$ \\
$\Xi_c(2790)^{0}$ & $1/2(?^?)$ &*** & $2791.8 \pm 3.3$ & $ < 12$ \\
$\Xi_c(2815)^{+}$ & $1/2(?^?)$ &*** & $2816.6 \pm 0.9$ & $ < 3.5$ &  $2845.2 \, (44.0)$\\
$\Xi_c(2815)^{0}$ & $1/2(?^?)$ &*** & $2819.6 \pm 1.2$ & $ < 6.5$ \\
$\Xi_c(2930)$ & $1/2(?^?)$ &* & $2931 \pm 6$ & $ 36 \pm 13$ \\
$\Xi_c(2980)^{+}$ & $1/2(?^?)$ &*** & $2971.4 \pm 3.3$ & $ 26 \pm 7$ \\
$\Xi_c(2980)^{0}$ & $1/2(?^?)$ &*** & $2968.0 \pm 2.6$ & $ 20 \pm 7$ \\
$\Xi_c(3055)$ & $1/2(?^?)$ &** & $3054.2 \pm 1.3$ & $ 17 \pm 13$ \\
$\Xi_c(3080)^{+}$ & $1/2(?^?)$ &*** & $3077.0 \pm 0.4$ & $ 5.8 \pm 1.0$ \\
$\Xi_c(3080)^{0}$ & $1/2(?^?)$ &*** & $3079.9 \pm 1.4$ & $ 5.6 \pm 2.2$ \\
$\Xi_c(3123)$ & $1/2(?^?)$ &* & $3122.9 \pm 1.3$ & $ 4 \pm 4$ \\
\hline
\end{tabular}
\end{center}
\caption{Summary of experimental data for baryon resonances in the different charm sectors as compiled in Ref.~\cite{Beringer:1900zz}. The lowest-lying $\Lambda_c$, $\Sigma_c$, $\Sigma^*_c$, $\Xi_c$, $\Xi^{\prime}_c$, $\Xi^*_c$, $\Omega_c$, $\Omega^*_c$, $\Xi_{cc}$, $\Xi^*_{cc}$, $\Omega_{cc}$, $\Omega^*_{cc}$ and $\Omega_{ccc}$ states are omitted. The mass and the width of the dynamically-generated states using the extended WT to SF with HQSS constraints are shown in the last column. The predicted values for $\Xi_c(2790)$ and $\Xi_c(2815)$ are the isospin-average states.}
\label{tab:exp}
\end{table}

The experimental status on charmed baryon states is summarized in Table~\ref{tab:exp}. Most of the states have been observed in the $C=1$ sector with zero strangeness. Some of these states can be identified with dynamically generated resonances~\footnote{Often we  refer to all poles generically as resonances, regardless of their concrete
  nature, since usually they can decay through other channels not included in  the model space.} using unitarized meson-baryon coupled-channel models. In the following we present the results using the extended WT interaction to SF with HQSS constraints as well as results from other unitarized coupled-channel schemes. This identification is made by comparing the PDG data~\cite{Beringer:1900zz} on these states with the mass, width and, most important, the coupling to the meson-baryon channels of our dynamically-generated poles. 

\subsection{\bf $ \bf C=1,2,3$ baryon states} 

We start by showing the results for the $C=1,2,3$ sectors, where the meson-baryon channels with $c \bar c$ pairs have been dropped out. We will consider in the last subsection the case when a $c \bar c$ is present, but only for the $C=0$ case.

In the $C=1,S=0, I=0, J=1/2$ case,  we obtain three states  in the $\Lambda_c$ sector, which come from the most attractive SU(6) $\times$ HQSS representations (irrep), denoted by ${\bf 15_{2,1}}$ and ${\bf 21_{2,1}}$. Here we use the notation $R_{2J_C+1,C}$, where {\bf R} is the SU(6) irrep label (for which we use the dimension) and $J_C$ is the spin carried by the quarks with charm.  The experimental $\Lambda_c(2595)$ resonance can be identified with the pole that we obtain around $2618.8\,\MeV$. The width  is, however, smaller than the experimental result since the dominant three-body decay channel $\Lambda_c \pi \pi$ \cite{Beringer:1900zz} is not included in our calculation. A second broad $\Lambda_c$ resonance at 2617.3~MeV is observed with a large coupling to the open channel $\Sigma_c \pi$, very close to $\Lambda_c(2595)$. This is the same two-pole pattern found in the charmless $I = 0, S = -1$ sector for the $\Lambda(1405)$ \cite{Jido:2003cb}. A third spin-$1/2$ $\Lambda_c$ resonance is seen around 2828~MeV  and cannot be assigned to any experimentally known resonance. With regard to spin-$3/2$ resonances, we find one located at $(2666.6 -i 26.8\,\MeV)$ coming from the attractive ${\bf 15_{2,1}}$ irrep. This resonance is assigned to the experimental $\Lambda_c(2625)$ and forms with the broad 2617.3 resonance a HQSS doublet. The $t$-channel vector-exchange model of Ref.~\cite{Hofmann:2006qx} also reported a similar resonance  at $2660\,\MeV$. The novelty of our calculations with respect to that one is that we obtain a non-negligible contribution from the vector meson-baryon channels to the generation of this resonance.

For $C=1, S=0, I=1, J=1/2$ ($\Sigma_c$ sector), three $\Sigma_c$ resonances are obtained from ${\bf 15_{2,1}}$ and ${\bf 21_{2,1}}$ irreps with masses 2571.5, 2622.7 and 2643.4~MeV and widths 0.8, 188.0 and 87.0~MeV, respectively. Those states are pure predictions of our model. Ref.~\cite{JimenezTejero:2009vq} predicts the existence of two resonances in this sector. However, only one of them can be identified to one of ours but with a strong vector meson-baryon component. Moreover, we predict two spin-$3/2$ $\Sigma_c$ resonances appearing from ${\bf 21_{2,1}}$ irrep. The first one, a bound state at 2568.4~MeV, is thought to be the charmed counterpart of the $\Sigma(1670)$. The second state at $2692.9 -i 33.5$ ~MeV has not a direct experimental comparison.

Our model generates six $\Xi_c$ states with $J=1/2$ and three with $J=3/2$  from ${\bf 15_{2,1}}$ and ${\bf 21_{2,1}}$ irreps. In this sector there are two negative-parity experimentally known resonances
that can be identified with some of our dynamically-generated states, namely experimental
$\Xi_c(2790)~J^P=1/2^-$ and
$\Xi_c(2815)~J^P=3/2^-$~ \cite{Beringer:1900zz}. 
The state $\Xi_c(2790)$ has a width of $\Gamma<12-15\,\MeV$ and it decays to
$\Xi_c' \pi$, with $\Xi_c' \rightarrow \Xi_c \gamma$. 
We assign it to the $2804.8 -i 10.3$~MeV state found in our model because of the large 
$\Xi'_c\pi$ coupling. A slight modification of the subtraction point can lower the
position of our resonance to $2790\,\MeV$ and most probably reduce its width as it will get
closer to the $\Xi_c' \pi$ channel. It could be also possible to
identify our pole at $2733\,\MeV$  with the experimental $\Xi_c(2790)$ state. In that case, one
would expect that if the resonance position gets closer to the physical mass
of $2790\,\MeV$, its width will increase and it will easily reach values of
the order of $10\,\MeV$. With regards to the $\Xi_c$ resonance with $J^P=3/2^-$, 
the full width of the experimental $\Xi_c$  is
expected to be less than $3.5\,\MeV$ for $\Xi_c^+(2815)$ and less than
$6.5\,\MeV$ for $\Xi_c^0(2815)$, and the decay modes are $\Xi_{c}^+ \pi^+
\pi^-$, $\Xi_{c}^0 \pi^+ \pi^-$. We obtain two resonances at $2819.7 -i 16.2\,\MeV$ and
$2845.2 -i 22.0\,\MeV$, respectively, that couple strongly to $\Xi_c^* \pi$, with
$\Xi_c^* \rightarrow \Xi_c \pi$. Allowing for this possible indirect
three-body decay channel, we might identify one of them to the experimental
result. This assignment is possible for the state at $2845.2\,\MeV$
if we slightly change the subtraction point, which will lower its
position and reduce its width as it gets closer to the threshold of the open $\Xi_c^* \pi$
channel. With this assignment, we find that the experimental $\Xi_c(2790)$ and $\Xi_c(2815)$ are HQSS partners.

Three $\Omega_c$ bound states are obtained from ${\bf 15_{2,1}}$ and ${\bf 21_{2,1}}$ irreps with masses $2810.9$, $2884.$5 and 
$2941.6$~MeV. There is no experimental information on those excited states. However, our
predictions can be compared to recent calculations of Refs.~\cite{Hofmann:2005sw,JimenezTejero:2009vq}.
In Ref.~\cite{JimenezTejero:2009vq} three $\Omega_c$ resonances are predicted, with masses higher than ours by approximately 100~MeV.
Further, we obtain two spin-$3/2$ bound states $\Omega_c$ from  the attractive ${\bf 21_{2,1}}$ irrep with masses $2814.3$ and
$2980.0$~MeV, which mainly couple to $\Xi D^*$ and $\Xi^* D^*$, and to $\Xi_c^* \bar K$,
respectively. As in the $J=1/2$ sector, no experimental information is
available here.

No experimental information is available in $C=2$ ($\Xi_cc$ and $\Omega_{cc}$) and $C=3$ ($\Omega_{ccc}$). Several states are generated dynamically in our model coming from the SU(6) $\times$ HQSS ${\bf 6_{1,2}}$ and  ${\bf 6_{3,2}}$ representations ($\Xi_{cc}$ and $\Omega_{cc}$ states), while from ${\bf 1_{2,3}}$ and  ${\bf 1_{4,3}}$ for  $\Omega_{ccc}$, all of them bound states in the $\Omega_{cc}$ and $\Omega_{ccc}$ sectors. For $\Omega_{ccc}$ the large separation from the closest threshold suggests that interaction mechanisms beyond $s$-wave could be relevant for the formation of the dynamically-generated states with $C=3$.

 \subsection{\bf $C=-1$ baryon states}

A very rich spectrum of exotic states is obtained in the $C = -1$ sector. In this case, only meson-baryon states with one $\bar c$ quark are considered. For $J = 1/2$ we find 6 SU(3) multiplets of resonances, while for $J=3/2$ our model generates 5 SU(3) multiplets and a SU(3) multiplet in $J=5/2$. 
The predictions of Refs.~\cite{Hofmann:2005sw,Hofmann:2006qx} within a $t$-channel vector exchange model and in Ref.~\cite{Wu:2004wg} in the Skyrme model are reproduced with the extended WT interaction. However, our states turn out to be about 100-200 MeV heavier. Using the extended WT interaction new states are also predicted because of the inclusion of vector mesons. 

Among the resonances we call the attention to a $C=-1$, $S = 0$, $I = 0$, $J=1/2$ state
generated by the $ \bar D N$ and $\bar D^* N$ coupled channel dynamics. This state appears bound by
only 1 MeV, and it is one of our more interesting predictions. Moreover, it appears as
a consequence of treating heavy pseudoscalars and heavy vector mesons on an equal
footing, as required by HQSS because no resonance would be generated unless $\bar D^* N$ channel is considered.

One of the poles we obtain for $J = 3/2$ has a mass close to 3100 MeV. There is one
experimental claim for an exotic state with $C = -1$ around this mass in Ref.~\cite{Aktas:2004qf}. The state
has been observed in the decay mode
\be
\Theta_{\bar C} \rightarrow \bar D^* N \rightarrow \bar D \pi N .
\label{decay}
\ee
Our dynamically generated state has other two possible decay channels induced by its
coupling to channels involving the $\Delta(1232)$ resonance, which is not a stable particle.
Thus, the anti-charmed resonance can decay to $\bar D$ or $\bar D^*$ plus a virtual $\Delta$, which subsequently
would decay into a $\pi N$ pair. So, the dynamically generated state has  another
two decay mechanisms apart from the one in Eq.~(\ref{decay}), namely,
\begin{eqnarray}
&&\Theta_{\bar C} \rightarrow \bar D \Delta \rightarrow \bar D \pi N \\
\label{decay2}
&&\Theta_{\bar C} \rightarrow \bar D^* \Delta \rightarrow \pi \bar D \pi N \ .
\label{decay3}
\end{eqnarray}
The decay in Eq.~(\ref{decay2}) has the same particles in the final state that in Eq.~(\ref{decay}), with the
difference that the pion in one case is coming from the decay of the $\bar D^*$ and therefore has
low momentum, while in the other channel it comes from a $\Delta$ and may have higher
momentum. The experimental search made in Ref.~\cite{Aktas:2004qf} looked only for pions in order to
reconstruct a $\bar D^*$ and may have missed the other events where the pion comes from a
$\Delta$.

\subsection{\bf Hidden-charm baryon states: the $C=0$ case}

In this section we analyze the hidden charm sector, that is, the dynamically generated states with $c \bar c$ pairs stemming from meson-baryon interactions. We focus on the $C=0$ case, so that  the resonances can be labeled as $N$- and $\Delta$-  states.  These states stem from the two attractive SU(6) $\times$ HQSS representations in the hidden-charm sector with $C=0$, i.e., $\bf{56_{2,0}}$ and $\bf{70_{2,0}}$ irreps.  Due to the presence of a $c \bar c$ pair, these $N$ and $\Delta$ states have masses around 4 GeV and no experimental counterparts for these states have been found yet.

We find three $N_{1/2}$ (the lower index indicates $J$), three $N_{3/2}$, 
and one $N_{5/2}$, with masses between 3918 and 4027 MeV. Some 
of the states are degenerated when the HQSS is unbroken, thus forming
the HQSS multiplets. Almost all found $N$ resonances are bound states.

Compared to other works, the $N$ resonances studied in the zero-range $t$-channel vector-meson exchange model of Refs.~\cite{Hofmann:2005sw,Hofmann:2006qx} are about 500 MeV lighter than those found in our model. The hidden-gauge
formalism predicts these masses to be about 400 MeV larger~\cite{Wu:2010vk,Wu:2012md,Xiao:2013yca}. However, this difference comes mostly from using a different renormalization prescription~\cite{Xiao:2013yca}. Our results are close to those predicted by the chiral interaction, studied in the constituent 
quark model of Ref.~\cite{Yuan:2012wz}, whereas the instanton-induced interaction
and color-magnetic interaction produce higher masses for the resonances~\cite{Yuan:2012wz}.

Further, we find three $\Delta_{1/2}$ and two $\Delta_{3/2}$ 
bound states. Two of them, $\Delta_{1/2}(4306)$ and $\Delta_{3/2}(4307)$,
which stem from the $\bf 56_{2,0}$ representation, appear as cusps in the scattering amplitude. 

\section{Open charm in dense nuclear matter}

The mass and width of the dynamically-generated states described in the previous section can be modified once these states are obtained in a dense medium. This is a direct consequence of the changes in matter experienced by the mesons and baryons that generate these states. The properties of open-charm mesons in matter have been object of a recent theoretical interest  due to the consequences for charmonium suppression, as observed at SPS energies by the NA50 collaboration \cite{Gonin:1996wn}. The change of the properties of D mesons in matter should modify the $J/\Psi$ absorption in a hot and dense nuclear medium and can provide an explanation for $J/\Psi$ suppression~\cite{Zhang:2000nc,Cassing:1997kw,Bratkovskaya:2003ux,Sibirtsev:1999jr}. Furthermore, there have been speculations about the existence of $D$ meson bound states in nuclei \cite{Tsushima:1998ru} resulting from a strong attractive potential felt by the $D$ mesons in nuclear matter \cite{Sibirtsev:1999js}.  

Several theoretical works have addressed the properties of open-charm mesons in matter. While some studies have been devoted to the properties of open charm with light mesons \cite{Fuchs:2004fh,Molina:2008nh}, over the past decade the interest has been also focused on open charm at finite baryonic density. A phenomenological estimate based on the quark-meson coupling (QMC) model \cite{Guichon:1987jp}, which is based on the exchange of $\omega$, $\rho$ and $\sigma$ mesons among the quarks in the meson/baryon bag, predicts an attractive $D^+$-nucleus potential at normal nuclear matter density ($\rho_0$) of the order of -140 MeV \cite{Sibirtsev:1999js}. The $D$-meson mass shift has also been studied using the QCD sum-rule (QSR) approach \cite{Hayashigaki:2000es,weise,Hilger:2008jg,Hilger:2011cq,Hilger:2012db}.  In these works the mass modification of the $D$-meson has a large contribution from the light quark condensates. A mass shift of -50 MeV at $\rho_0$  has been suggested \cite{Hayashigaki:2000es}. A second analysis based on QSR, however, predicts only a splitting of $D^+$ and $D^-$ masses of 60 MeV at $\rho_0$ because the uncertainties to which the mass shift is subject at the level of the unknown $DN$ coupling to the sector of charm baryons and pions \cite{weise}.  Recent results on QSR rules for open-charm mesons can be found in \cite{Hilger:2008jg,Hilger:2011cq,Hilger:2012db}.  On the other hand, the mass modification of the $D$-meson has been also addressed using a chiral effective model in hot and dense matter \cite{Mishra:2003se,Kumar:2010gb,Kumar:2011ff} within the mean field or relativistic Hartree Fock approaches, and strong mass shifts were obtained. Whereas similar results to the previous works on QMC model or QSR are obtained at $T=0$ with the interaction Lagrangian of chiral perturbation theory, a larger mass drop of $\sim$ -200 MeV at $T=0$ is observed when a SU(4) effective model is use. At finite temperature, the attraction is, though, reduced.

Nevertheless, in all these investigations the full spectral features (mass and width) of the open-charm mesons in dense nuclear matter have not been considered. Thus, a self-consistent unitarized coupled-channel approach is required.

\subsection{Self-consistent unitarized meson-baryon coupled-channel schemes}

The spectral features of open-charm mesons and, hence, the self-energy in nuclear matter are obtained by incorporating the corresponding medium modifications in the effective open charm-nucleon interactions.  This amounts for solving a unitarized coupled-channel equation similar to Eq.~(\ref{eq:bse}), where the intermediate meson-baryon propagators contain different sources of density dependence. This procedure requires the implementation of self-consistency for open charm. Indeed, the effective open charm-nucleon interactions in matter are used to calculate the open-charm spectral features (masses and widths). These features modify the intermediate open charm-nucleon propagators which enter in the calculation of the effective interactions in matter and, thus, the solution requires a self-consistent treatment.

\begin{figure}
\begin{center}
\includegraphics[width=0.3\textwidth,height=0.35\textwidth]{tolos1.eps}
\includegraphics[width=0.69\textwidth,height=0.6\textwidth]{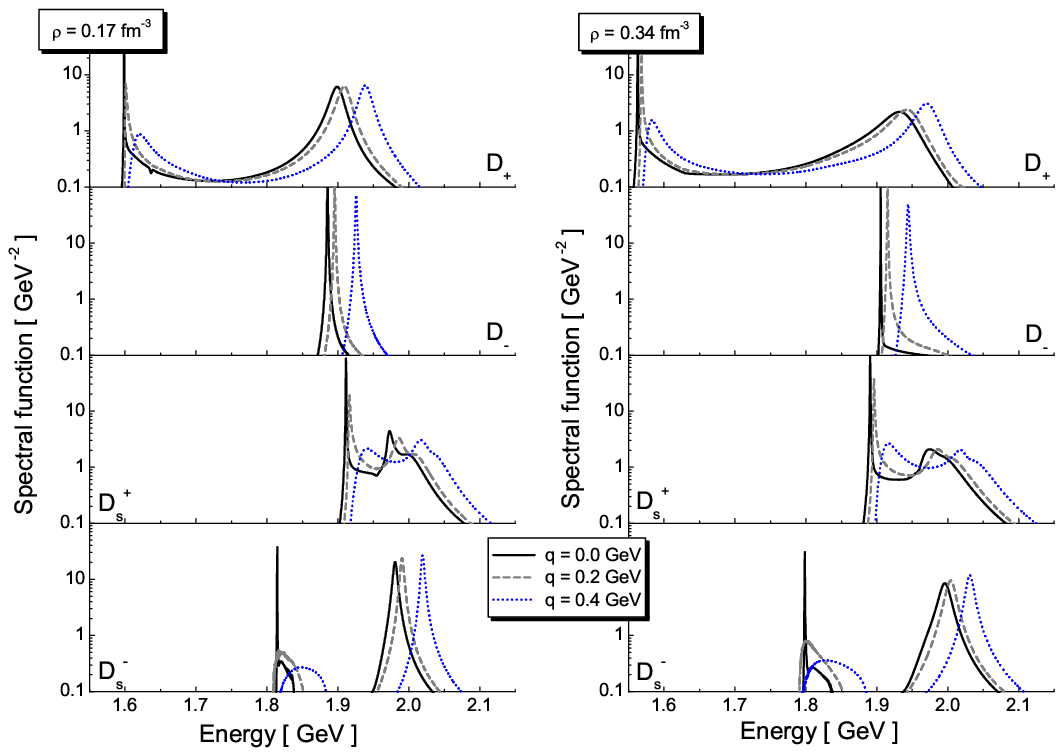}
\caption{ Left figure (taken from Ref.~\cite{Tolos:2005ft}): The $D$ spectral function at zero momentum for different densities at $T=120$ MeV together with the $ \rho_0,T=0$ case within the self-consistent coupled-channel approach that assumes a separable potential for the bare meson-baryon interaction of Ref.~\cite{Tolos:2005ft} . Right figure (taken from Ref.~\cite{Lutz:2005vx}): The $D$, $\bar D$, $D_s$ and $\bar D_s$ spectral functions at different momenta for $\rho_0$ and 2$\rho_0$ within the $t$-channel vector-meson exchange model in matter of Ref.~\cite{Lutz:2005vx} . }
\label{fig:others1}
\end{center}
\end{figure}

\begin{figure}
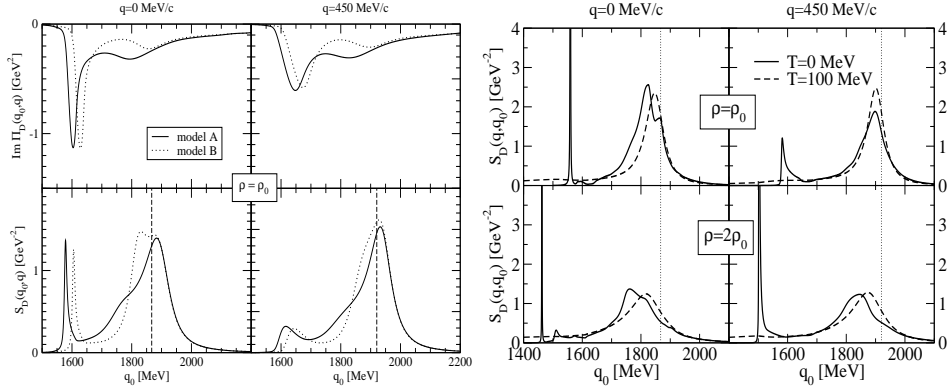

\begin{center}
\includegraphics[width=0.49\textwidth,height=0.405\textwidth]{Fig_7_spec_cut_2.eps}
\includegraphics[width=0.49\textwidth,height=0.405\textwidth]{Fig6.eps}
\caption{Left figure (taken from Ref.~\cite{Mizutani:2006vq}): The $D$ meson self-energy (upper panels) and spectral function (lower panels) at $\rho=\rho_0$ for two momenta within the $t$-channel vector-meson exchange model in the zero-range approximation of Ref.~\cite{Mizutani:2006vq} . Two different approaches are depicted. Right figure (taken from Ref.~\cite{JimenezTejero:2011fc}): The $D$ meson spectral function for $\rho_0$ and $2 \rho_0$, for different momenta and temperatures within the model that uses a $t$-channel vector meson exchange approach beyond the zero-range approximation of Ref.~\cite{JimenezTejero:2011fc} .}
\label{fig:others2}
\end{center}
\end{figure}

The $D$ meson mass and width  were first obtained in the exploratory work of Ref.~\cite{Tolos:2004yg} within a self-consistent coupled-channel approach assuming a separable potential for the bare meson-baryon interaction. Finite temperature effects in the $D$ meson-nucleon interaction were incorporated later on in Ref.~\cite{Tolos:2005ft}. The spectral function for $D$ mesons can be seen in the left-hand plot of Fig.~\ref{fig:others1}. Within the $t$-channel vector-meson exchange model of Refs.~\cite{Hofmann:2005sw,Hofmann:2006qx}, the spectral features of the $D, \bar D$ and $D_s, \bar D_s$ mesons in nuclear matter were analyzed in the zero-range limit \cite{Lutz:2005vx}. These results were revisited in Refs.~\cite{Mizutani:2006vq,Tolos:2007vh}. The results are shown in the right-hand plot of Fig.~\ref{fig:others1} for the first model while in the left-hand plot of Fig.~\ref{fig:others2} for the revisited one.  More recently, the spectral features of $D, \bar D$ and $D_s, \bar D_s$ mesons have been addressed taking for the interaction kernel a $t$-channel vector meson exchange model beyond the zero-range approximation  \cite{JimenezTejero:2011fc} (see right-hand plot of Fig.~\ref{fig:others2} for the $D$ meson spectral function).

These models in dense matter, however, are not consistent with HQSS, as heavy vector mesons and pseudoscalars are not treated on equal footing. In the following, we address the implementation of HQSS constraints in the determination of the spectral features of open-charm mesons.


\subsection{Implementation of heavy-quark spin symmetry}

As seen in Sec.~\ref{su8wt}, a spin-flavor symmetric model for four flavors that implements HQSS constraints has been developed in Refs.~\cite{GarciaRecio:2008dp,Gamermann:2010zz,Romanets:2012hm,Garcia-Recio:2013gaa,GarciaRecio:2012db} , similarly to the SU(6) approach in the light sector of Refs.~\cite{GarciaRecio:2005hy, GarciaRecio:2006wb, Toki:2007ab, GarciaRecio:2010ki,Gamermann:2011mq,Garcia-Recio:2013uva} . This model implements HQSS \cite{Isgur:1989vq} and dynamically generates resonances with negative parity in all the isospin, spin, strange and charm sectors  that one can form from an s-wave interaction between pseudoscalar and vector meson multiplets with $1/2^+$ and $3/2^+$ baryons. 

Within this model, the properties of open-charm mesons in nuclear matter can be obtained by incorporating the corresponding medium modifications in the effective open charm-nucleon interactions. One of the sources of density dependence comes from the Pauli principle acting on the nucleons, which prevents the scattering to already occupied nucleon states. Another source is related to the change of the properties of mesons and baryons in the intermediate states of the coupled-channel structure due to the interaction with nucleons of the Fermi sea.
Those changes are implemented by using the in-medium meson-baryon propagators instead of the corresponding free ones. 

The renormalized (finite) loop function in free space of Eq.~(\ref{eq:loop}) can be rewritten as \cite{Tolos:2009nn}
\begin{eqnarray}
G^0(\sqrt{s}) &=& {\rm i} 2 M \int \frac{d^4 q}{(2 \pi)^4}
\left( D^0_{\cal B}(P-q)~D^0_{\cal M}(q)
-D^0_{\cal B}(\bar{P}-q)~D^0_{\cal M}(q) \right) \ ,
\label{eq:g0sustr}
\end{eqnarray}
where $D^0_{\cal B (\cal M)}$ is the free baryon (meson) propagator and
with $P$ and $\bar P$ defined such that $P^2=s,~ {\bar P}^2 =
(\mu)^2$, with $\mu$ the subtraction point. For simplicity, the $C$,$S$,$I$ and $J$ indices have been omitted.

In matter, the medium changes appear as a correction to the renormalized (finite) loop function of free space \cite{Tolos:2009nn}
\begin{eqnarray}
G_\rho(P) &=& G^0(\sqrt{s}) + \delta G_\rho(P) \ , \nonumber \\
\label{eq:defGrhob}
\delta G_\rho(P) &=& {\rm i}
2 M \int \frac{d^4 q}{(2 \pi)^4} \left ( D^\rho_{\cal
B}(P-q)~D^\rho_{\cal M}(q)-D^0_{\cal B}(P-q)~D^0_{\cal M}(q) \right ) ,
\end{eqnarray} 
 where $D^\rho_{\cal B(\cal M)}$ are the baryon (meson) propagator calculated at finite
density $\rho$.

We can now solve the on-shell Bethe-Salpeter equation in nuclear matter for 
the in-medium amplitudes using the spin-flavor extended WT interaction that respects HQSS constraints:
\begin{eqnarray}
{T^\rho}^{(CSIJ)} &=& \frac{V^{CSIJ}}{1-
V^{CSIJ}\,{G_\rho}^{CSIJ}}\,\ .
 \label{eq:scat-rho}
\end{eqnarray}
In particular, the self-energies of open charm $D$ and $\bar D$ and their HQSS partners $D^*$ and $\bar D^*$ can be calculated simultaneously in a self-consistent manner. These self-energies are obtained by
integrating the effective interaction in matter, ${T^{\rho}_{D(D^*)N}}$, over the nucleon Fermi sea, 

\begin{eqnarray}
\Pi_{D(\bar D)} (q_0,{\vec q}\,) &=&
\int \frac{d^3p}{(2\pi)^3}\, n(\vec{p}\,)  \,  [\, {T^{\rho}_{D(\bar D) N}}^{(I=0,J=1/2)} +
3 \, {T^{\rho}_{D(\bar D) N}}^{(I=1,J=1/2)}\, ] \nonumber \\ 
\Pi_{D^*(\bar D^*) }(q_0,\vec{q}\,) &=& \int \frac{d^3p}{(2\pi)^3} \, n(\vec{p}\,) \,
\, \left [~ \frac{1}{3} \, {T^{\rho}_{D^*(\bar D^*)N}}^{(I=0,J=1/2)}+
{T^{\rho}_{D^*(\bar D^*)N}}^{(I=1,J=1/2)}+ \right . \nonumber \\
&&  \left . \frac{2}{3} \,
{T^{\rho}_{D^*(\bar D^*)N}}^{(I=0,J=3/2)}+ 2 \,
{T^{\rho}_{D^*(\bar D^*)N}}^{(I=1,J=3/2)}\right ] \  ,
\label{eq:selfd}
\end{eqnarray}
\noindent
where $P_0=q_0+E_N(\vec{p},T)$ and $\vec{P}=\vec{q}+\vec{p}$ are
the total energy and momentum of the meson-nucleon pair in the nuclear
matter rest frame, and ($q_0$,$\vec{q}\,$) and ($E_N$,$\vec{p}$\,) stand  for
the energy and momentum of the meson and nucleon, respectively, in this
frame. The self-energy is determined self-consistently since it is obtained from the
in-medium amplitude which contains the meson-baryon loop function, and this quantity itself
is a function of the self-energy. Then, the meson spectral function  reads
\begin{eqnarray}
S_{M}(q_0,{\vec q}\,)= -\frac{1}{\pi}\frac{{\rm Im}\, \Pi_{M}(q_0,\vec{q}\,)}{\mid
q_0^2-\vec{q}\,^2-m_{M}^2- \Pi_{M}(q_0,\vec{q}\,) \mid^2 } \ .
\label{eq:spec}
\end{eqnarray}
for $M=D (\bar D) ,\bar D(\bar D^*)$.

\begin{figure}
\begin{center}
\includegraphics[width=0.7\textwidth,height=0.5\textwidth]{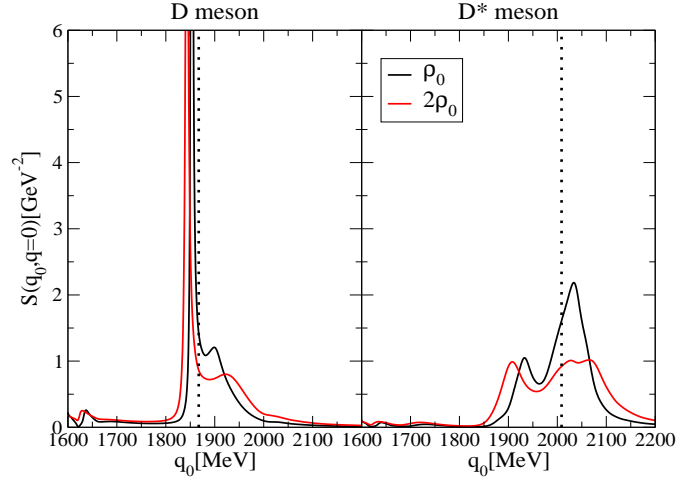}
\caption{The $D$ and $D^*$ spectral functions  for $\rho_0$ and $2 \rho_0$ at $\vec{q}=0$ MeV/c. Taken from Ref.~\cite{Tolos:2009nn} .}
\label{fig:spec}
\end{center}
\end{figure}

The $D$ and $D^*$ spectral functions for $\rho_0$ and $2 \rho_0$ ($\rho_0=0.17 {\rm fm^{-3}}$) at zero momentum  are displayed in Fig.~\ref{fig:spec}.  Any dynamically-generated resonance which strongly couples to $DN$ and/or $D^*N$ channels will appear as a resonant-hole  state in the corresponding spectral function for energies around the mass of the resonance once the nucleon mass is subtracted. The peak of the $D$ meson spectral function mixes strongly with $\Sigma_c(2823)N^{-1}$ and $\Sigma_c(2868)N^{-1}$ states, where we have denote $N^{-1}$ as hole. These $\Sigma_c$ resonances are predictions of our model with no experimental confirmation yet. Note that these resonances have higher masses than those described in Sec.~\ref{dyn}. These states were seen in the wider energy range explored in Ref.~\cite{GarciaRecio:2008dp} but do not come from the most attractive SU(6) $\times$ HQSS representations. The $\Lambda_c(2595)N^{-1}$ is clearly visible in the low-energy tail.  With regard to the $D^*$ meson, the $D^*$ spectral function incorporates the $J=3/2$ resonances, and the peak of the spectral function fully mixes with the  $\Sigma_c(2902)N^{-1}$ and $\Lambda_c(2941)N^{-1}$ states. The  $\Sigma_c(2902)$ can be identified with the experimental $\Sigma_c(2800)$\cite{Beringer:1900zz}, as seen in Ref.~\cite{GarciaRecio:2008dp} while $\Lambda_c(2941)$ has no experimental correspondence yet. Again these states are not reported in Sec.~\ref{dyn} as they do not come from the most attractive SU(6) $\times$ HQSS representations. For both $D$ and $D^*$ mesons, the $Y_c(=\Lambda_c^{(*)},\Sigma_c^{(*)})N^{-1}$ modes tend to smear out and the spectral functions broaden with increasing phase space, as seen before in the $t \rightarrow 0$ vector-meson exchange model of Ref.~\cite{Mizutani:2006vq}. Compared to the previous results for the $D$ meson spectral function of Refs.~\cite{Tolos:2004yg,Lutz:2005vx,Mizutani:2006vq,JimenezTejero:2011fc}, we find that all of them generate the $\Lambda_c(2595) N^{-1}$ for energies around 1600 MeV, however with different strength. Moreover, while the works of Refs.~\cite{Tolos:2004yg,Lutz:2005vx,Mizutani:2006vq,JimenezTejero:2011fc} also generate the experimental $\Sigma_c(2800)$ resonance, though, in different positions, the model based on the extended WT to SF with HQSS constraints contains not only this resonance but also a richer spectrum of states.

The optical potential for $D$ and $\bar D$  meson is shown in  Fig.~\ref{fig:pot}. Defined as
\begin{equation}
  V_{D(\bar D)}(\rho,E) = \frac{
  \Pi_{D(\bar D)}(q^0=m_{D(\bar D)}+E,\vec{q}=0,~\rho)}{2 m_{D(\bar D)}},
\label{eq:UdepE}
\end{equation}
where $E=q^0-m_{D(\bar D)}$ is the $D$ or $\bar D$ energy excluding its mass, and $\Pi_{D(\bar D)}$ the meson self-energy, we observe in both cases a strong energy dependence close to the open-charm meson mass. In the case of the $D$ meson, this is due to the mixing of the peak  of the spectral function with the  $\Sigma_c(2823)N^{-1}$ and $\Sigma_c(2868)N^{-1}$ states. For the $\bar D$ meson, the presence of a bound state at 2805 MeV, discussed in Sec.~\ref{dyn} and found in Ref.~\cite{Gamermann:2010zz} almost at $\bar D N$ threshold, makes the potential also strongly energy dependent. This is in contrast to the $t \rightarrow 0$ vector-meson exchange model of Ref.~\cite{Mizutani:2006vq}, also displayed in Fig.~\ref{fig:pot} for $\rho_0$ with the SU(4) label.

\section{Charmed mesons in nuclei}

In this final section we aim at reviewing possible experimental signatures of the in-medium modified open-charm properties. One possible scenario is the so-called $D$ and $\bar D$-meson bound states in nuclei.

$D$ and $\bar D$-meson bound states in $^{208}$Pb were predicted in Ref.~\cite{Tsushima:1998ru} relying upon an attractive  $D$ and $\bar D$ -meson potential in the nuclear medium. This potential was obtained within a QMC model \cite{Sibirtsev:1999js}. The experimental observation of those bound states, though, might be problematic since, even if there are bound states, their widths could be very large compared to the separation of the levels. This is indeed the case for the potential derived from a SU(4) $t$-vector meson exchange model for $D$-mesons \cite{Mizutani:2006vq,Tolos:2007vh}.


\begin{figure}
\begin{center}
\includegraphics[width=0.8\textwidth,height=0.5\textwidth]{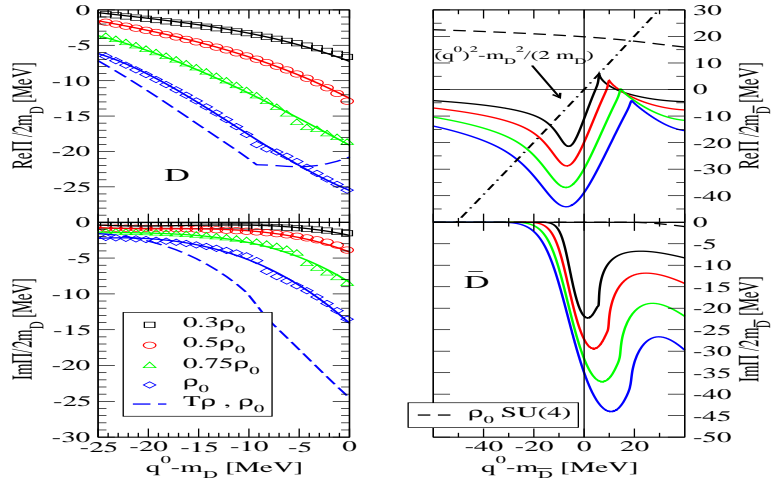}
\caption{ The $D$ and $\bar D$ optical potential at $\vec{q}=0$ MeV/c for different densities. Taken from Refs.~\cite{GarciaRecio:2010vt,GarciaRecio:2011xt} .}
\label{fig:pot}
\end{center}
\end{figure}


We look for $D$- and $\bar D$- nucleus bound states \cite{GarciaRecio:2010vt,GarciaRecio:2011xt} by  solving the
 Schr\"odinger equation in the local density approximation:
 \begin{equation}
  \left[ -\frac{{\bm \nabla}^2}{2 m_{\rm red}} +  V_{\rm{coul}}(r)
+  V_{\rm{opt}}(r) \right] \Psi \,
  = (-B-i \Gamma /2) \Psi
.
\label{eq:SchE}
\end{equation}
In this equation, $B$ is the binding energy ($B>0$), $\Gamma$ the width
of the bound state and $m_{\rm red}$ is the $D (\bar D)$-nucleus reduced
mass. $V_{\rm coul}(r)$ is the Coulomb potential only for $D^-$ mesons
including the nucleus finite size and the Uehling vacuum
polarization. We use the energy dependent optical potential $V_{\rm{opt}}(r)=V((\rho(r),E)$, with $V(\rho,E)$ defined in Eq.~(\ref{eq:UdepE}) and shown in Fig.~\ref{fig:pot}.  We apply the local density approximation in order to determine the relation between the density $\rho$ and the radius $r$ for each given nucleus density profile. Because the electromagnetic interaction is introduced by means of
the minimal coupling prescription (to be consistent with gauge invariance and
electric charge conservation), $V_{\rm coul}(r)$ must be introduced wherever
the energy is present. So the energy dependent optical potential is applied with argument $q^0=m-B-V_{\rm coul}(r)$.

\begin{figure}[t]
\begin{center}
\includegraphics[width=0.42\textwidth,height=0.48\textwidth,angle=-90]{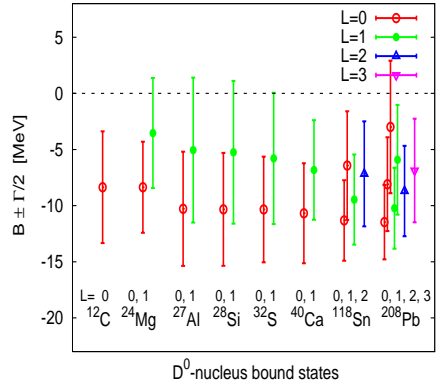}
\caption{$D^0$-nucleus bound states. Taken from Ref.~\cite{GarciaRecio:2010vt} . \label{fig:d0}}
\end{center}
\end{figure}

The question is whether $D$ and/or $\bar D$  will be bound in nuclei. We start by discussing  $D$ mesons in nuclei.  We observe that the $D^0$-nucleus states are weakly bound (see Fig.~\ref{fig:d0}), in contrast to previous results using the QMC model \cite{Tsushima:1998ru}. Moreover, those states have significant widths \cite{GarciaRecio:2010vt}, in particular  for $^{208}$Pb \cite{Tsushima:1998ru}. The best chances for observation of bound states are in the region of $^{24}\mbox{Mg}$, provided an orbital angular momentum separation can be done, where there is only one $s-$ bound state and its half width is about a factor of two smaller than the binding energy. Only $D^0$-nucleus bound states are possible since the Coulomb interaction prevents the formation of observable bound states for $D^+$ mesons.

\begin{figure}[t]
\begin{center}
\includegraphics[width=0.48\textwidth,height=0.45\textwidth]{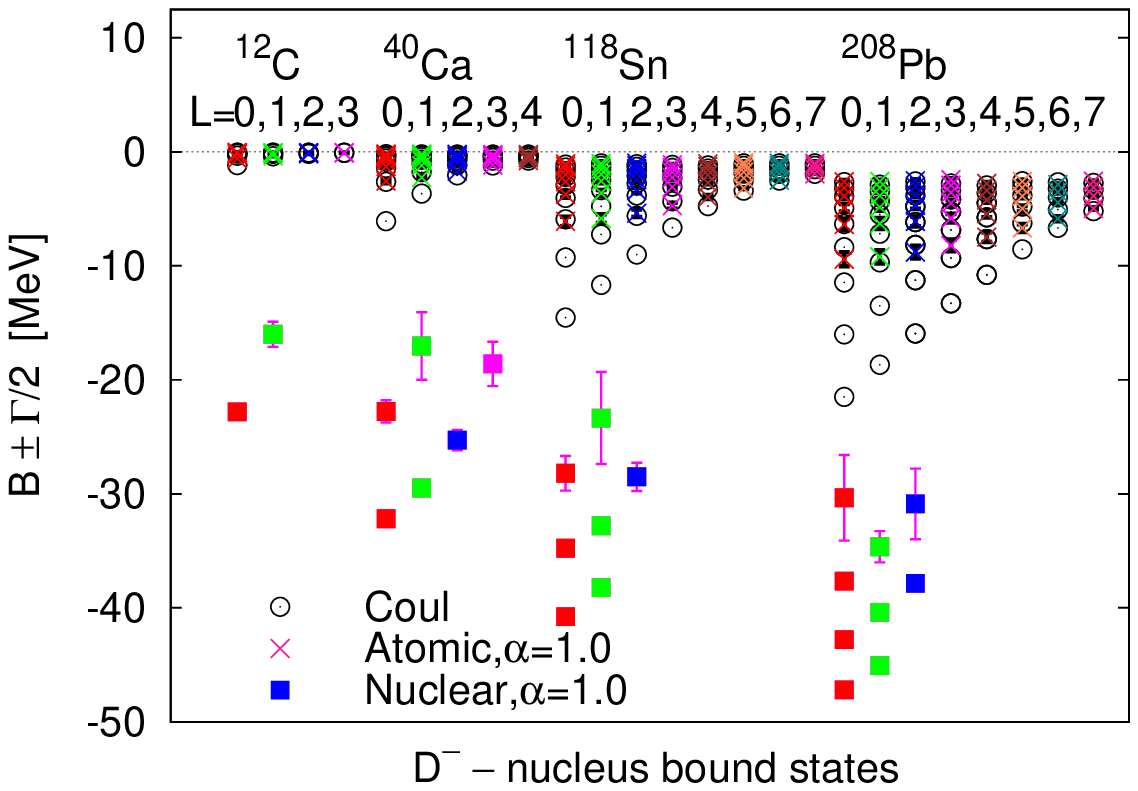}
\hfill
\includegraphics[width=0.48\textwidth,height=0.45\textwidth]{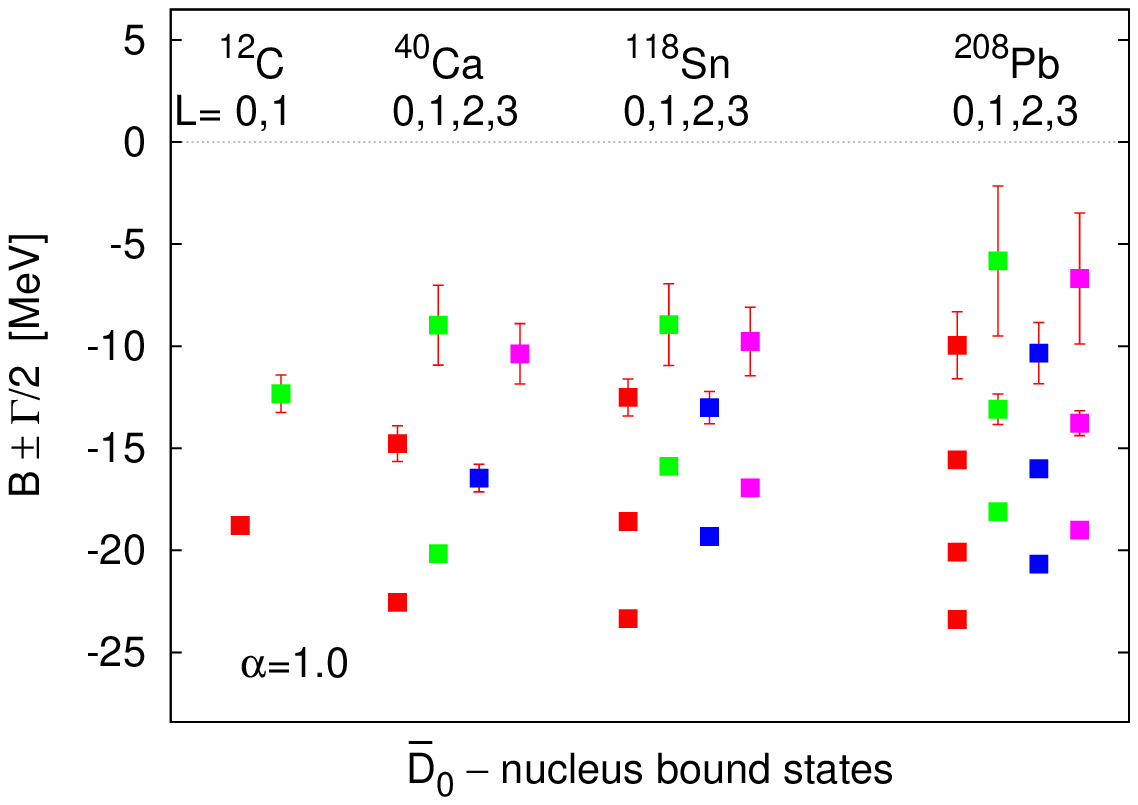}
\caption{$D^-$ and $\bar D^0$- nucleus bound states. Taken from Ref.~\cite{GarciaRecio:2011xt} \label{fig:dmd0}  .}
\end{center}
\end{figure}

Apropos of  $\bar D$-mesic nuclei, not only $D^-$ but also $\bar{D}^0$ bind in nuclei as seen in Fig.~\ref{fig:dmd0}. The spectrum contains states of atomic and of nuclear types for all nuclei for $D^-$  while, as expected, only nuclear states are present for $\bar{D}^0$ in nuclei. Compared to the pure Coulomb levels, the atomic states are less bound. The nuclear ones are more bound and may present a sizable width \cite{GarciaRecio:2011xt}. Moreover, nuclear states only exist for low angular momenta.

In Ref.~\cite{Bayar:2012dd} a bound state of $DNN$ is found with a large binding energy
(bigger than 100 MeV) and a relatively small width. The existence of $DN$
(namely the $\Lambda_c(2595)$) and $DNN$ bound states, suggests that
$D$-nucleus deeply bound states might also exist for larger nuclei.  This
would be consistent with the optical potential displayed in Fig.~3 of Ref.~\cite{Tolos:2009nn},
where a strongly attractive region is displayed around 200 MeV below
threshold.  In the results reported in this review, however, only  the region near threshold has been
considered.

\begin{figure}
\begin{center}
\includegraphics[width=0.5\textwidth]{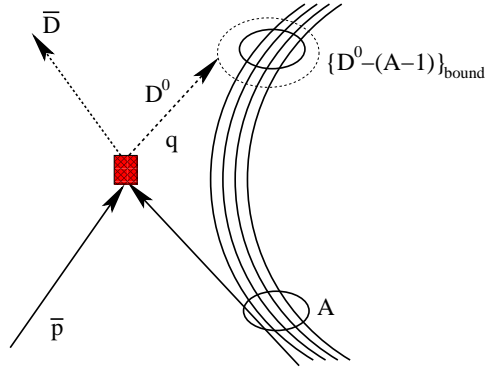}
\caption{Possible production of $D^0$-mesic nuclei with an antiproton beam\label{fig:mechanism}. Taken from Ref.~\cite{GarciaRecio:2010vt} . }
  \end{center}
\end{figure}

The information on bound states is very valuable to gain some knowledge on the charmed meson-nucleus interaction, which is of interest for PANDA at FAIR. The experimental detection of $D$ and $\bar D$-meson bound states is, though, a difficult task. For example, reactions with antiprotons on nuclei (Fig.~\ref{fig:mechanism}) for obtaining $D^0$-nucleus states might have a very low production rate, as seen in Ref.~\cite{GarciaRecio:2010vt}. Reactions but with proton beams, although difficult, seem more likely to trap a $D^0$ in nuclei \cite{GarciaRecio:2010vt}.

\section{Summary}

We have reviewed the properties of charmed mesons in dense matter and nuclei. We have started by discussing unitarized coupled-channel approaches in free space, paying a special attention to the model that takes, as bare interaction, effective Lagrangians that respect chiral symmetry in the light sector while HQSS in the heavy one. Within this scheme,  several resonances have been generated dynamically by the $s$-wave interaction between pseudoscalar and vector meson multiplets with $1/2^+$ and $3/2^+$ baryons. Some of them have been identified with experimental results, such as the $\Lambda_c(2595)$, the $\Lambda_c(2625)$, the $\Xi_c(2790)$ or the $\Xi_c(2815)$ states, the last two  forming a HQSS doublet. Moreover, we have overviewed the properties of open-charm mesons in nuclear matter by studying different self-consistent coupled-channel approaches and, in particular, the one which explicitely respects HQSS constraints. Finally, we have analyzed the experimental signatures of the in-medium properties of the open-charm mesons, such as the possible formation of charmed mesic nuclei in the context of FAIR.

\section*{Acknowledgements}
The author warmly thanks C. Garcia-Recio, J. Nieves, O. Romanets and L.L. Salcedo for the careful reading of the manuscript. This research was supported  by Ministerio de Ciencia e Innovaci\'on under contract FPA2010-16963, the EU HadronPhysics3 project (Grant Agreement No. 283286), the Ramon y Cajal Research Programme, and the FP7-
PEOPLE-2011-CIG under Contract No. PCIG09-GA-2011-291679.

\end{document}